\documentclass[aps,pra,twocolumn,amsmath,amssymb,nofootinbib,showpacs,superscriptaddress]{revtex4-1}
\usepackage[english]{babel}
\usepackage{latexsym}
\usepackage{graphics}
\usepackage{graphicx}
\usepackage{epsfig}
\usepackage{color}
\usepackage{bm}
\usepackage{amsmath}
\usepackage{amssymb}
\usepackage{amsthm}
\usepackage{dcolumn}
\usepackage{bm}
\usepackage{float}
\usepackage{hyperref}
\usepackage{color}
\usepackage{epstopdf}
\usepackage{cleveref}
\usepackage[svgnames]{xcolor}
\hypersetup{hidelinks,colorlinks=true,allcolors=DarkBlue}

\renewcommand{\sb}{\mathbf{s}}
\newcommand{\sbx}{\mathbf{s_x}}
\newcommand{\sby}{\mathbf{s_y}}
\newcommand{\xb}{\mathbf{x}}
\newcommand{\xeb}{\mathbf{x}_{\mathrm{ext}}}
\newcommand{\xd}{\mathbf{x}_{\mathrm{dec}}}
\newcommand{\yb}{\mathbf{y}}
\newcommand{\yeb}{\mathbf{y}_{\mathrm{ext}}}
\newcommand{\eb}{\mathbf{e}}
\newcommand{\Hb}{\mathbf{H}}
\newcommand{\hb}{h_\mathrm{b}}
\newcommand{\rb}{\mathbf{r}}
\newcommand{\zb}{\mathbf{z}}
\newcommand{\epse}{\epsilon_\mathrm{est}}
\newcommand{\ed}{\eb_\mathrm{dec} }

\newcommand{\qe}{q_\mathrm{est}}

\newcommand{\Mb}{\mathbf{M}}
\newcommand{\LLR}{\mathrm{LLR}}

\newcommand{\Nm}{\mathcal{N}}
\newcommand{\Mm}{\mathcal{M}}

\newcommand{\fst}{f_{\mathrm{start}}}

\newcommand{\modt}{({\rm mod}~2)}

\begin{document}

\preprint{APS/123-QED}

\title{Symmetric blind information reconciliation for quantum key distribution}

\author{E.O. Kiktenko}\email{e.kiktenko@rqc.ru}
\affiliation{Russian Quantum Center, Skolkovo, Moscow 143025, Russia}
\affiliation{QApp, Skolkovo, Moscow 143025, Russia}
\affiliation{Steklov Mathematical Institute of Russian Academy of Sciences, Moscow 119991, Russia}

\author{A.S. Trushechkin}
\affiliation{Steklov Mathematical Institute of Russian Academy of Sciences, Moscow 119991, Russia}

\author{C.C.W. Lim}\email{elelimc@nus.edu.sg}
\affiliation{Department of Electrical and Computer Engineering, National University of Singapore, 117583, Singapore}
\affiliation{Quantum Information Science Group, Computational Sciences and Engineering Division, Oak Ridge National Laboratory, Oak Ridge, Tennessee 37831-6418, USA}

\author{Y.V. Kurochkin}
\affiliation{Russian Quantum Center, Skolkovo, Moscow 143025, Russia}

\author{A.K. Fedorov}\email{akf@rqc.ru}
\affiliation{Russian Quantum Center, Skolkovo, Moscow 143025, Russia}
\affiliation{QApp, Skolkovo, Moscow 143025, Russia}

\date{\today}
\begin{abstract}
Quantum key distribution (QKD) is a quantum-proof key-exchange scheme which is fast approaching the communication industry. 
An essential component in QKD is the information reconciliation step, which is used for correcting the quantum-channel noise errors. 
The recently suggested blind-reconciliation technique, based on low-density parity-check codes (LDPC), 
offers remarkable prospectives for efficient information reconciliation without an {\it a priori} rate estimation. 
We suggest an improvement of blind-information-reconciliation protocol promoting a significant increase in the efficiency of the procedure and reducing its interactivity. 
The proposed technique is based on introducing symmetry in operations of parties, and the consideration of results of unsuccessful belief-propagation decodings.
\end{abstract}
\maketitle

\section{Introduction}

Quantum key distribution (QKD)  is the art of distributing provably-secure cryptographic keys in an insecure communications network~\cite{BB84,BB843,Gisin,Scarani}. 
Unlike conventional cryptography, the security of QKD is based on the laws of quantum physics and thus is guaranteed to be secure against any unforeseen technological and algorithmic developments, {\it e.g.}, quantum computing.~For this reason, QKD has attracted an enormous amount of interest since its discovery, and is now one of the most widely studied research field in quantum information science.~In fact, provably secure commercial QKD systems are now available at retail~\cite{Market}. 

A QKD system is carried out in two consecutive phases, namely, a quantum key establishment 
phase and a classical postprocessing phase~\cite{Renner}.~In the first phase, the users first create an unprocessed (raw) key pair by performing local measurements on quantum signals which are exchanged via an untrusted quantum channel.~At this point, the pair of raw keys are weakly correlated---due to noise in the quantum channel---and are partially secure.~To correct the errors and remove the adversary's information about the raw key pair, the users run an information reconciliation step and a privacy amplification step.~The former requires the user to exchange a certain amount of public information about the key pair, which is then compensated for in the privacy amplification step.~Finally, after the classical post-processing phase, the users are left with a correct and secure key pair~(for details, see Refs.~\cite{Renner,Gisin2,SECOQC,Fedorov}).

\begin{figure*}
	\includegraphics[width=0.725\linewidth]{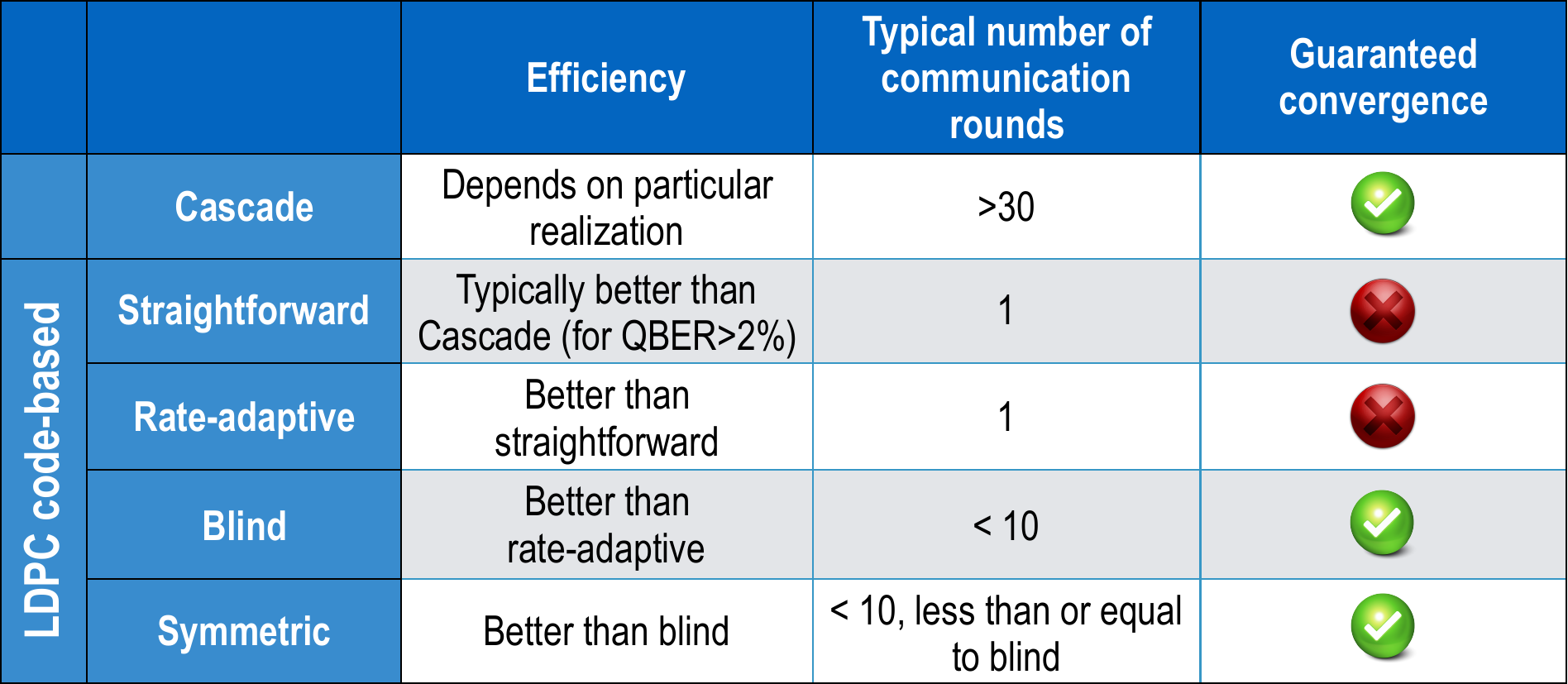}
	\caption{Comparison of three parameters (efficiency, typical number of communication rounds, and convergence guarantee) 
	for different approaches to information reconciliation for QKD systems: 
	the cascade method~\cite{Cascade2_rev}, 
	straightforward implementation of LDPC codes~\cite{LDPC2_poly}, 
	rate-adaptive implementation of LDPC codes~\cite{LDPC4_adapt1}, 
	blind reconciliation implementation of LDPC codes~\cite{LDPC5_blind}, 
	and our proposed solution (symmetric blind information reconciliation). 
	We note that the typical number of communication rounds for blind and symmetric blind protocols 
	is considered in the assumption that both the protocols are running until a convergence of syndrome decoding.}
	\label{fig:comparison}
\end{figure*}

It is clear that information reconciliation is an important step in QKD, for it is necessary to correct the errors introduced by the quantum channel (or the adversary).
~In practice, 
the information reconciliation step is typically implemented using an iterative method known as Cascade~\cite{Cascade1_orig}.~The Cascade 
method is based on the random shuffling and the dichotomic search of discrepancies based on announcements of sub-block parities via communication over the authenticated channel.~A number 
of possible improvements have been proposed, but most of them are very expensive in terms of communication~\cite{Cascade2_rev}.
That is, despite the fact that different sub-blocks can be treated in parallel~\cite{Cascade3_par}, 
the Cascade method is still a highly interactive algorithm as the dichotomic search requires multiple rounds of communication between users.

Interactivity of Cascade-based information reconciliation procedure can cost significant amount of authentication resources together with time delays and the workload in QKD systems.~Another popular information reconciliation scheme is forward error correction with LDPC codes~\cite{LDPC_orig1,LDPC_orig2}, which uses a single message containing a syndrome calculated for a particular block of a sifted key~\cite{LDPC1_prop,LDPC2_poly,LDPC3_something,LDPC4_adapt1,LDPC4_adapt2}.
However, this scheme could fail and penalize the secret key throughput due to its inability to perform the syndrome-decoding procedure.
Such failures appear if the syndrome decoding, based on an iterative belief-propagation algorithm, does not converge in the predefined number of iterations 
(\emph{e.g.} it could be caused by an inappropriate choice of the code rate relative to the actual number of discrepancies in raw keys).

The problem with convergence differentiates the traditional LDPC code-based error correction methods~\cite{LDPC_orig1,LDPC_orig2} from the Cascade, 
where the dichotomic search is performed as long as all of the sub-blocks in all of the shuffling rounds contain odd numbers of errors.~Then Cascade can be considered as a \emph{guaranteed-convergence} method (see Fig.~\ref{fig:comparison}).
It is important to note that guaranteed convergence does not imply \emph{guaranteed reconciliation}.
In the case of Cascade, some of sub-blocks still can contain positive numbers of undetected errors after implementation of the reconciliation procedure~\cite{Cascade2_rev}.
The analogous problem remains for all the LDPC-code-based reconciliation protocols, where belief-propagation decoding sometimes can converge to an inappropriate codeword.
In order to solve this problem, an additional step of verification with universal hashing is usually considered~\cite{Gisin2,ChiHang}.

Therefore, an important task for optimizing the workflow of QKD is to provide a regime with guaranteed convergence of the information reconciliation scheme, but without significant expenditure of authentication and time resources.~This can be achieved by combining the key advantages of the aforementioned schemes and by introducing some interactivity into error correction with LDPC codes.~This technique is known as blind information reconciliation~\cite{LDPC4_adapt1,LDPC5_blind,LDPC6_blindrep} and can operate without an \emph{a priori} estimation of the quantum bit error rate (QBER).

In this work, we demonstrate further improvements of error correction combining LDPC codes and interactivity. 
We show that the use of interactivity --- by introducing symmetry in operations of parties and the consideration of results of unsuccessful belief-propagation decodings --- 
allows one to perform an efficient and convergence guaranteed information reconciliation procedure.~For practical QKD parameters, simulation results show an average of about 
${10}\%$ improvement in efficiency and an average of about ${30}\%$ improvement in the number of information requests.~We refer to our proposed method as the symmetric blind information reconciliation.
For a comparison of the proposed information reconciliation procedure with existing solutions, see Fig.~\ref{fig:comparison}.

The paper is organized as follows. 
In Sec.~\ref{sec:basic}, we explain concepts of the information reconciliation procedure.
In Sec.~\ref{sec:blind}, we present an improvement of blind information reconciliation with LDPC codes.
We summarize our results and consider an illustrative example in Sec.~\ref{sec:conclusion}. 

\section{Basic concepts of error correction}\label{sec:basic}

The goal of the information reconciliation procedure is to correct the errors between Alice's and Bob's raw keys 
by disclosing some key information over a public (authenticated) channel.~Each bit value of the Bob's string is a result of a transmission of the corresponding bit from the Alice's string through a binary symmetric channel (BSC).
The crossover probability $q$ of the channel is also known as the QBER.

One of the ways to perform error correction is to use a LDPC code which is a linear code with a sparse $m\times n$ binary parity-check matrix~\cite{LDPC_orig1,LDPC_orig2}.
Alice multiplies the parity-check matrix by a block of the raw key of length $n$ to obtain a syndrome of length $m$, which is then sent to Bob.~Then, Bob performs a syndrome-decoding operation on his side using his raw key, 
the same sparse matrix, and an estimated level of QBER, which comes from the preceding procedures.

In the best-case scenario, the syndrome-decoding procedure outputs the same key as is on Alice's side.
Nevertheless, there is still a probability of an undetected frame error.
To ensure that the error-correction procedure is performed properly, an additional stage of error verification is applied~\cite{Gisin2,ChiHang}.
It can be done using a universal hashing technique~\cite{Carter1,Carter2}, which guarantees correctness with a probability depending on the length of the hash code.
There is also a possibility that the syndrome decoding based on belief-propagation procedure does not converge in the specified number of iterations. 
Then the parties have to discard the processed blocks of the raw key and go to the next ones.

An important figure of merit for a reconciliation protocol is its efficiency $f$. 
It is given by the redundancy of disclosed information to the theoretical limit necessary for successful reconciliation~\cite{SlWolf}.
For a given BSC it is characterized by the Shannon binary entropy of the QBER~\cite{BinaryEntropy}: 
\begin{equation} \label{eq:ffix}
	\hb(q)=-q\log_2q-(1-q)\log_2(1-q).
\end{equation}
Thus, the efficiency of the considered information reconciliation with LDPC code can be represented as
\begin{equation} \label{eq:ffix}
	f = \frac{m}{n\hb(q)}=\frac{1-R}{\hb(q)},
\end{equation}
where $R=1-m/n$ is the rate of a given LDPC code.
The importance of the efficiency $f$ is based on the fact that the value of disclosed information has to be removed from the key in the stage of privacy amplification.
We also note that the efficiency larger than unity does not guarantee successful decoding.
In fact, it depends on the specific parity-check matrix, the maximal number of iteration in decoding procedure and other factors.

\subsection{Rate-adaptive scheme}

The straightforward implementation of the LDPC error correction suffers from the following drawback. 
The efficiency parameter $f$ is fixed by the dimension of the parity-check matrix and the current level of the QBER, according to Eq.~(\ref{eq:ffix}).
A naive way to perform information reconciliation with the desired efficiency is to choose or construct another parity-check matrix with a new rate, {\it i.e.}, an $m/n$ ratio.

Two elegant ways known as shortening and puncturing have been proposed to adjust the rate of the LDPC code to the desirable efficiency by modification of encoding and decoding vectors rather than through the parity check matrix~\cite{LDPC4_adapt1,LDPC4_adapt2}.
The main idea is to perform syndrome coding and decoding with extended keys of length $n$ obtained from the original raw keys of length $n-s-p$ by padding them with $s$ shortened and $p$ punctured bits.~The shortened symbols are the ones which have values exactly known by Alice and Bob, as well as by the adversary.~The values of punctured bits come from true random number generators (TRNGs), independently of the both sides. 
In this way, the shortened (punctured) bits serve for lowering (raising) the average discrepancy between the extended keys.
The positions for shortened and punctured bits could be chosen using a synchronized pseudo-random number generator (PRNG), or depending on a particular parity-check matrix (for example, via the untainted-puncturing method~\cite{UntP}).

After the construction of the extended keys, the parties perform information reconciliation in the same way as discussed above.
The difference is that in the case of a successful decoding, Bob excludes shortened and punctured bits from the result of the decoding procedure to obtain a corrected version of his raw key.
The efficiency of the described scheme is defined in the following form:
\begin{equation} \label{eq:fad}
	f = \frac{m-p}{\left[n-p-s\right]\hb(q)}.
\end{equation}
Thus, the artificial reduction (increase) of discrepancies between extended keys by shortened (punctured) 
bits allows one to implement fine-tuning of efficiency in order to keep a trade-off between a probability of failure for belief-propagation decoding and information leakage.

\subsection{Blind reconciliation}

The above scheme implies a single message sent from Alice to Bob only.~This is a crucial advantage over the cascading method, which is highly interactive~\cite{Cascade2_rev}.~However, the cascading method demonstrates rather good efficiency, particularly at low values of the QBER~\cite{Cascade2_rev,LDPC2_poly}.~Also, cascade methods do not suffer from the inability to perform the error correction, {\it i.e.}, it always converges to some result.~Therefore, the cascading method is widely used as an important benchmark for comparison of information reconciliation protocols~\cite{LDPC6_blindrep}.

To combine ``the best of both worlds'' by linking interactivity with the LDPC codes, a blind information reconciliation technique was suggested~\cite{LDPC4_adapt1,LDPC5_blind,LDPC6_blindrep}.
Its title comes from the fact that it can operate without an \emph{a priori} estimation of the QBER 
(a rough estimation of the QBER for the belief-propagation decoding one can be obtained directly from the syndromes~\cite{Estimating}).  
Blind reconciliation is based on the hybrid automatic repeat request technique~\cite{HARQ} with the LDPC codes with an essential presence of punctured symbols.
The crucial difference is that, in the case of a decoding failure, parties try to implement the decoding procedure again by turning a number of punctured symbols into shortened ones instead of discarding their blocks.~The values of these bits are transferred via the classical channel after a corresponding request of Bob.~Once the punctured bits are exhausted it is possible to continue revealing additional information with sifted key bits.
The efficiency of the procedure after $n_\mathrm{add}$ number of additional communication rounds is given by~\cite{LDPC5_blind}
\begin{equation}\label{eq:fadapt}
	f = \frac{m-p_0+n_\mathrm{add}d}{\left[n-p_0-s_0\right]\hb(q)},
\end{equation}
where $s_0$ and $p_0$ are the initial numbers of punctured and shortened bits, and $d$ is the number of disclosed bits in each additional round of blind reconciliation.
The meaning of expression~\eqref{eq:fadapt} relative to with expression~\eqref{eq:fad} is as follows: if the decoding procedure according to the rate-adaptive scheme with efficiency~\eqref{eq:fad} does not converge, 
then the parties increase $f$ in each additional communication round of the blind reconciliation to increase the probability of convergence.

The main advantage of blind reconciliation over the rate-adaptive scheme is that it allows one to adjust the efficiency to the actual error ratio, 
which can significantly fluctuate around the average QBER.
In Refs.~\cite{LDPC5_blind,LDPC6_blindrep} it was shown that the gradual disclosing of information can notably lower the mean value of $f$ together with frame error rate (FER).
These are benefits obtained at the price of introducing additional interactivity (see Fig.~\ref{fig:comparison}).

\section{Symmetric blind reconciliation}\label{sec:blind}

We suggest an improvement of blind information reconciliation with LDPC codes.
The proposed technique allows one to overcome the drawbacks of the aforementioned information reconciliation schemes 
by providing guaranteed belief propagation-based decoding with decreased information leakage and a decreased number of communication rounds.~Our approach is based on applying
information reconciliation with LDPC codes in a symmetric way. 
Specifically, it consists of the following general steps 
(a detailed description of the procedure is given in Methods).
First, in analogy to the rate-adaptive scheme, the parties choose the numbers and positions of the shortened and punctured bits and extend their blocks of raw keys.
Second, both Alice and Bob compute the syndrome of their extended raw keys and share them with each other.
Then they perform belief-propagation decoding. 
In a successful case, one party --- say Bob --- corrects the errors, and the procedure proceeds to the verification stage. 
In failed case, the parties exchange the values of a fixed number of bits having maximal uncertainty according to the log-likelihood ratio (LLR).  
After that, Alice and Bob repeat the belief-propagation decoding procedure with the updated list of shortened and punctured positions. 
In this respect, the proposed symmetric blind reconciliation is similar to the standard blind reconciliation.
The alternative ingredient is that the positions of additionally disclosed bits come not from the punctured positions but are decidedly indicated by an unsuccessful belief-propagation decoding algorithm.~This removes the restrictions on a number of additionally disclosed bits and also makes it possible to perform interactive LDPC code-based reconciliation even in the absence of punctured bits.
The latter scenario allows the adjustment of current sets of LDPC codes to a broad range of QBER values (see Appendix A).
We also note that, in contrast to the standard blind reconciliation protocol, 
where the parties use two consecutive messages in each of the communication rounds (the request and the corresponding answer), 
in the symmetric blind reconciliation the messages between parties are transferred simultaneously.

The convergence of the proposed method is formally guaranteed by the fact that, in the worst-case scenario, the parties reveal the entire extended key.~Clearly, in this case, the block will be useless for the secret key distillation.~In practice, the convergence takes place after a relatively small number of additional communication rounds.
The efficiency of the suggested method as of blind reconciliation is given by Eq. \eqref{eq:fadapt}.

The security analysis for rate-adaptive information reconciliation was considered in detail in Ref.~\cite{LDPC_security}.
It turns out that the symmetric blind reconciliation due to its inherent symmetry has a lot in common with the Cascade method, and the security analysis, 
developed for a class of adaptive symmetric error-correction methods (including the Cascade method) given in Ref.~\cite{LO_decoupling}, is applicable to our proposed scheme as well.~We refer the reader to Appendix C for details.

In order to demonstrate the improvements on the efficiency of the information reconciliation procedure, we perform a numerical simulation. 
Specifically, we compare the proposed procedure to the standard blind reconciliation, as in the most progressive LDPC based method for information reconciliation in the QKD systems.
We use a set of four standard LDPC codes~\cite{Standard} with the rates 
\begin{equation}\label{eq:codeset}
	\mathcal{R} = \{5/6, 3/4, 2/3, 1/2\},
\end{equation} 
with the block length fixed to $n=1944$.
For each of these codes, we obtain a list of bit positions according to the untainted-puncturing technique~\cite{UntP} containing $p_{\max}=$ 154, 221, 295 and 433 symbols, respectively.
These codes are currently used in industrial QKD systems~\cite{Gisin2,Fedorov}.

We simulate standard blind and symmetric blind reconciliation procedures with the absence of initially shortened bits and $p_{\max}$ 
initially punctured bits for a range of QBER values from $1\%$ up to $10.5\%$ (typical range for BB84 implementations).
In addition, we fix the FER to less than $10\%$.

The number of bits to be disclosed in each additional round of the procedure is chosen according to a particular code rate $R$ and the heuristic expression
\begin{equation}\label{eq:d}
	d(R) = \lceil n\cdot(0.0280-0.02R) \cdot \alpha \rceil ,
\end{equation}
where $n$ is the block length, $\alpha$ is the auxiliary parameter, and $\lceil\dots\rceil$ is the standard ceiling operation.
This expression was constructed in order to keep the mean number of additional communication rounds approximately on the same level for all the considered values of the QBER.
The larger the value of the parameter $\alpha$, the wider the step over the possible values of the efficiency $f$, 
for which the syndrome decoding process is tried [see Eq.~\eqref{eq:fadapt}], and the lower the number of iterations are required to perform reconciliation.
The pay for a small amount of communication rounds is that the resulting value of $f$ becomes higher, than in the case with a shorter step $d$.
Therefore, the parameter $\alpha$ allows us to realize a trade-off between the mean number of additional communication rounds and the mean efficiency of the information reconciliation.

\begin{figure*}
	\includegraphics[width=0.8\linewidth]{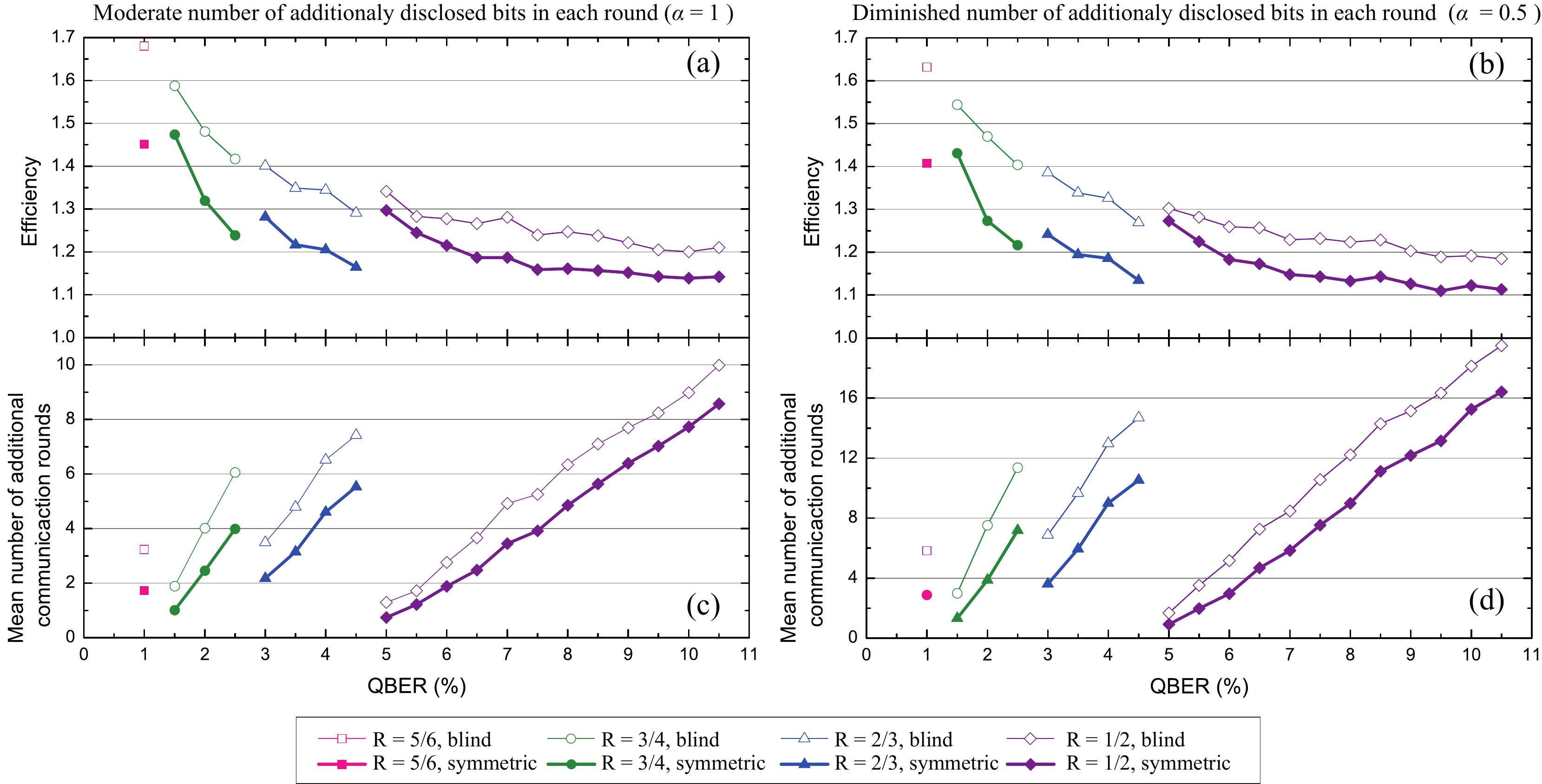}
	\vskip -2mm
	\caption
	{
	Comparison of the standard blind and symmetric blind information reconciliation protocols for two modes of disclosed bit number calculation~(\ref{eq:d}): 
	$\alpha=1$ (left column) and $\alpha=0.5$ (right column).
	Thin lines and empty symbols stand for blind information reconciliation, bold lines and filled symbols stand for the suggested approach (symmetric blind information reconciliation).
	In (a) and (b) the efficiencies, {\it i.e.} ratios of disclosed information to theoretical limit, are shown as functions of the QBER for the four standard LDPC codes~\cite{Standard} 
	with block length $n=1944$.
	In (c) and (d) the mean numbers of communication rounds are shown as function of the QBER.
	Changes of the codes (corresponding to different $R$) were performed according to the requirement that the probability of convergence for the blind reconciliation using only the punctured bits is larger than $90\%$. The convergence probability for the symmetric blind reconciliation is always $100\%$
	}
	\label{fig:blind_vs_sym}
\end{figure*}

The simulation results for $\alpha=1$ and $0.5$ are presented in Fig.~\ref{fig:blind_vs_sym}.
First, one can see that symmetric reconciliation improves both efficiencies $f$.~This comes from the fact that the decoding procedure in the symmetric scheme has a faster convergence rate.~Moreover, it requires a smaller number of additional communication rounds.

From these data, we identify an average of ${10}\%$ improvement in the efficiency 
($10.4\%$ for $\alpha = 1$, and $11.4\%$ for $\alpha=0.5$) and an average of ${30}\%$ improvement in the number of information requests ($28\%$ for $\alpha=1$, and $33\%$ for $\alpha=0.5$).
Moreover, the scheme does not suffer from the frame errors coming from unsuccessful belief-propagation decodings.

\begin{figure*}
	\includegraphics[width=0.62\linewidth]{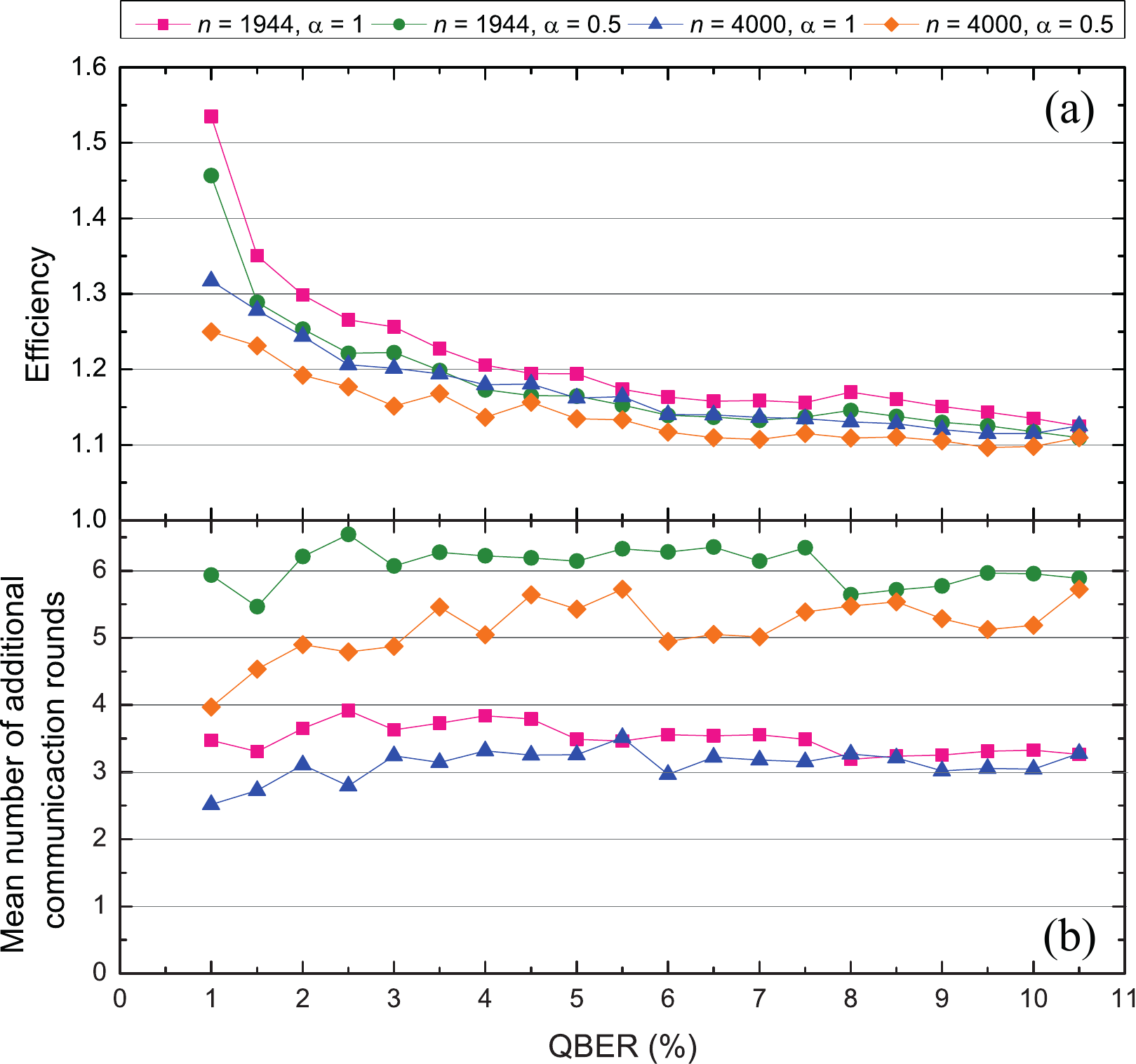}
	\vskip -3mm
	\caption
	{
	In (a) the efficiency of rate-adaptive symmetric blind reconciliation for two sets of codes and two modes of disclosed bit number is shown as a function of QBER. 
	In (b) mean numbers of additional communication round of rate-adaptive symmetric blind reconciliation for two sets of codes and two modes of disclosed bit number  
	is shown as function of QBER.
	The first set of codes is the standard one with block length $n=1944$~\cite{Standard} and the second set of codes consists of nine codes with rates~(\ref{eq:codeset2})
	and block length $n=4000$.
	The disclosed bit number chose according to~\eqref{eq:d} with $\alpha=1$ (moderate) and $\alpha=0.5$ (diminished).
	}
	\label{fig:full_qber}
\end{figure*}

Next, we compare two sets of codes in the rate-adaptive regime under the assumption that the level of the QBER is known.
The first set of codes is the previously considered one with rates~(\ref{eq:codeset}) and block length fixed to $n=1944$.
The second set of codes has rates in the range
\begin{equation}\label{eq:codeset2}
	\mathcal{R}'=\{0.5,0.55,\ldots,0.9\}
\end{equation} 
with the block length fixed to $n{=}4000$.
It is constructed with the use of the improved edge-growth algorithm~\cite{IPEG} with the degree distribution polynomials given by Ref.~\cite{LDPC2_poly}.
The initial numbers of shortened and punctured bits are chosen to obtain initial decoding efficiency $\fst=1$ (see Appendix B).
For each code we also construct a set of untainted-puncturing positions~\cite{UntP} and use them if possible.

The results of simulation for two sets of codes and two values of $\alpha$ (0.5 and 1) are presented in Fig.~\ref{fig:full_qber}.
It is clearly seen that the use of codes with block length $n=1944$ and $\alpha=0.5$ gives roughly the same efficiency as the codes with the block length $n=4000$ and $\alpha=1$.
This observation suggests that the symmetric blind reconciliation procedure is able to perform a trade-off between the number of required communication rounds and information leakage.

\section{Discussion and conclusion}\label{sec:conclusion}

In this paper, we propose an approach which significantly improves the blind information reconciliation technique --- 
the most progressive LDPC codes-based method for information reconciliation in the QKD systems. 
The gain comes from employing information from unsuccessful decodings and making the whole information reconciliation process in a symmetric form.

Specifically, propose disclosing a number of bits with the positions corresponding to maximal uncertainty of the values upon finishing the decoding procedure rather than certain bits in the punctured positions.
We note that the shortcoming of the presented method is that it occupies computational resources on the both sides 
and make it impossible to parallelize two opposite one-way information reconciliation processes. 
The ability of symmetric blind reconciliation to obtain rather low values of efficiency with short-length codes 
is expected to realize an efficient throughput with hardware-implemented syndrome decoding.

We note that short-length LDPC codes have been used to show the results of our method. 
The fact is that a small block length leads to high fluctuations in the actual number of discrepancies in raw keys even in the case of a constant QBER.
In turn, these fluctuations are crucial for successful belief-propagation decoding.
The feature of blind reconciliation is that it can treat fluctuations by disclosing an adequate amount of information via public channel.

The suggested method of information reconciliation can essentially be used for LDPC codes with large block lengths (say, $10^4$ or $10^5$).
In the case of an adjustment to a proper level of initial efficiency, it can be used for the complete elimination of belief-propagation decoding failures via a relatively rare request of additional bit values.
Nevertheless, these requests could appear to be very useful in the case of fluctuations of the QBER and especially in the case where error estimation is performed after the error correction (similar to that in Ref.~\cite{Gisin2}).

In order to evaluate the performance of our proposed scheme in the context of industrial QKD systems, we consider an illustrative example based on the results of Ref.~\cite{Gisin2}.
In this particular setup, the information reconciliation was performed with a straightforward implementation of the standard $n=1944$ LDPC code~\cite{Standard} 
with $R=3/4$ using QBER $q\approx 1.9\%$.
According to the results presented in Fig.~\ref{fig:full_qber}, 
an implementation of symmetric blind reconciliation may lead to a decrease of efficiency down to $f\approx 1.3$ with approximately six additional communication rounds.
It provides a $10\%$ increase in the secure key rate 
(we note that Cascade implementation of the information reconciliation procedure in the same conditions requires about $50$ communication rounds~\cite{Cascade2_rev}).

Moreover, in this QKD system, an estimated level of QBER is calculated via the comparison of a number of key blocks before and after error correction 
(unverified blocks are conservatively assumed to have a $50\%$ error rate).
Verification errors, resulting from unsuccessful belief-propagation decodings and convergences to improper vectors, 
leading to an overly pessimistic estimation of the QBER: 
$q_\mathrm{est}\approx3.4\%$.~Thus, the suggested approach opens a way for QBER estimation in a more accurate way along with a more economical utilization of generated raw keys.

Our source code for a proof-of-principle realization of the symmetric blind information reconciliation procedure 
for Python 2.7 is freely available under the GNU general public license (GPL)~\cite{Code}.~A proof-of-principle realization of the suggested post-processing procedure is also available~\cite{Code2}.

Finally we would like to emphasize that in the current contribution we conside the information ordering, inherent in the QKD scheme: 
(i) the sifted quantum keys are generated using consistent preparation and measurements of events, 
(ii) the information reconciliation for removing discrepancies in the sifted keys is applied, 
(iii) the privacy amplification, based on the estimated quantum-channel parameters and information leakage in the information reconciliation step, is performed for obtaining the pair of secret identical keys.
Then, this pair of keys can be used for information-theoretically secure information transfer with one-time pad encryption (Vernam cipher) or other applications~\cite{Alleaume}.
The above scheme differs from the another important protocol of quantum cryptography, namely quantum secure direct communication (QSDC), 
where the information is directly transmitted through the quantum channel without a preliminary key generation stage (\emph{e.g.} see Ref.~\cite{QSDC1} for the first theoretical proposal, 
and Ref.~\cite{QSDC2,QSDC3} for recent experimental demonstrations).
In the case of QSDC, there is also a need in the error correction, but usually it is more appropriate to perform it by means of some noisy-channel coding techniques: 
that is, by adding redundancy to the transmitting message, rather than by employing public discussion after the message has been transmitted.  

\section*{Acknowledgments}
We thank N. Gisin for a number of useful discussions and valuable comments.~We thank J. Mart\'inez-Mateo for numerous important remarks and comments helping us to improve the manuscript. 
We acknowledge our fruitful discussions with N. Pozhar, M. Anufriev, D. Kronberg, and D. Elkouss.~We thank M. Galchenkova for invaluable input in the initial phase of the project. 
The research leading to these results has received funding from Russian Science Foundation under project 17-71-20146.~C.C.W.L. acknowledges support from ORNL laboratory directed research and development (LDRD) program, 
the U.S.~Department of Energy Cybersecurity for Energy Delivery Systems (CEDS) program program under contract M614000329.

\setcounter{equation}{0}
\renewcommand{\theequation}{A\arabic{equation}}

\section*{Appendix A: Workflow of the symmetric blind reconciliation}

\begin{figure*}
\includegraphics[width=1\linewidth]{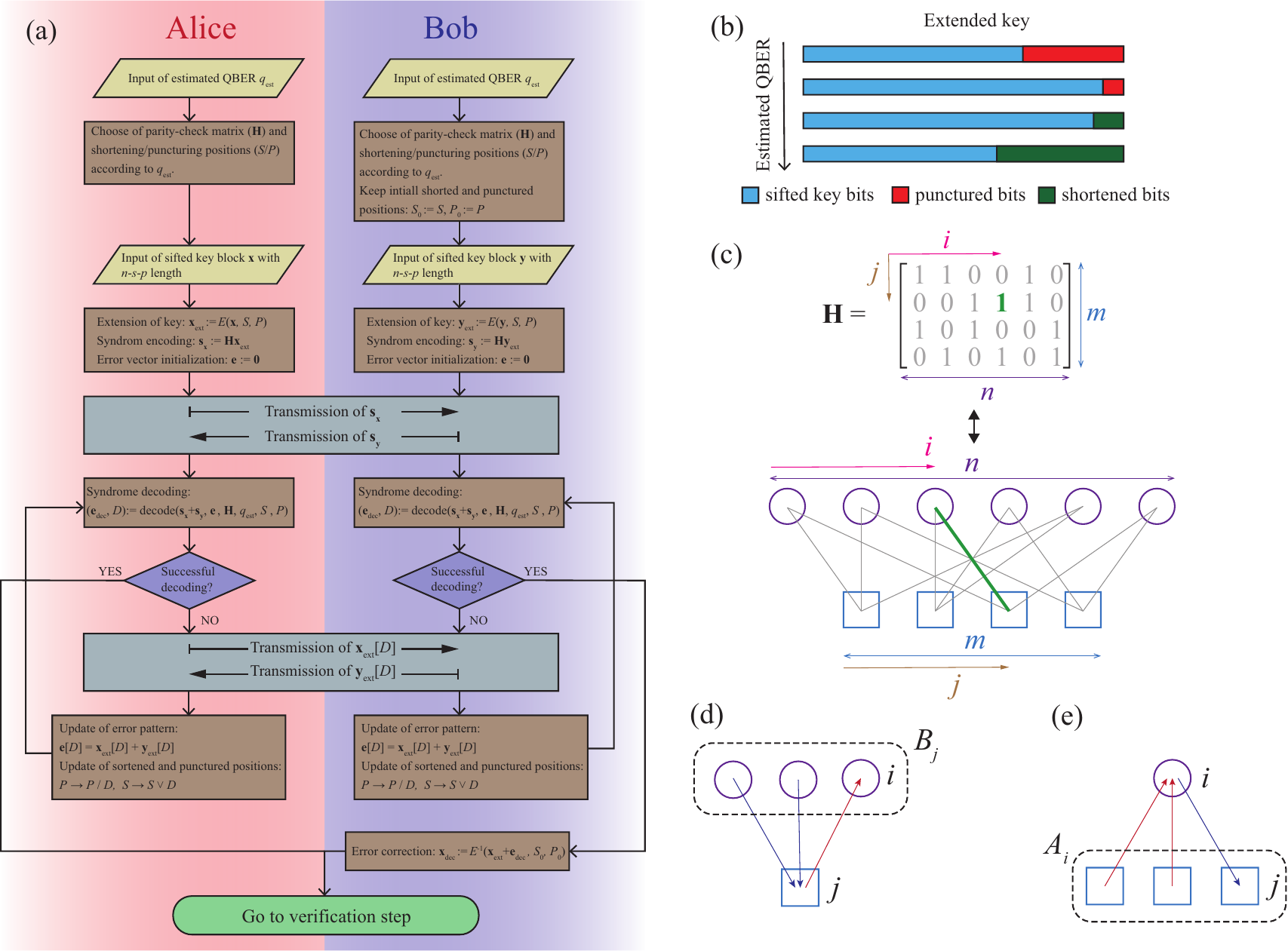}
\vskip -3mm
\caption
	{
	(a) The block scheme of a symmetric blind-reconciliation-procedure workflow is presented. Note that all summations are assumed to be performed by modulo 2. 
	(b) Visual representation of choosing numbers for shortened and punctured symbols in an extending key according to an estimated level of QBER for a particular code is shown. 
	(c) The correspondence between parity-check matrix and the bipartite Tanner graph is presented. 
	(d),(e) The principles of construction of messages from check node to symbol node and symbol node to check node are shown.
	}
\label{fig:methods}
\end{figure*}

Here, we give a detailed description of the workflow of the proposed symmetric blind reconciliation.
The general scheme is presented in Fig.~\ref{fig:methods}(a).

First, we assume that Alice and Bob have an estimated value of QBER $\qe$, which comes from the preceding error estimation step or previous rounds of post-processing procedure.
Parties start with choosing the optimal code from a set (pool) of available LDPC codes according to $\qe$ and desired starting efficiency $\fst$ (in all of our discussions, it is set to unity).
For each code, specified by its $m\times n$ parity matrix $\Hb$ (with $m<n$), 
parties calculate the number of shortened ($s$) or punctured $(p)$ symbols that is required to obtain desired efficiency $\fst$ from the nonadaptable efficiency $f_0=m/\left[n\;\hb(\qe)\right]$ as follows:
\begin{equation}\label{eq:sp}
\begin{aligned}
	& p=\lfloor {(m- n\hb(\qe)\fst)}/{(1-\hb(\qe)\fst)} \rfloor, \\ 
	& s=0
\end{aligned}
\end{equation}
for $f_0>f_{\mathrm{start}}$, and
\begin{equation}\label{eq:sp2}
\begin{aligned}
	& s= \lceil n - m/\hb(\qe)\fst\rceil, \\ 
	& p=0
\end{aligned}
\end{equation}
for $f_0<f_{\mathrm{start}}$.

The particular code among the set is then chosen in such a way that it has the maximal number of raw key bits in the extended key.
We note that, in our approach, we use only shortened {\it or} punctured bits to obtain the desired efficiency $\fst$ [see also Fig.~\ref{fig:methods}(b)]. 
This method is quite different from the commonly used approach~\cite{LDPC4_adapt1,LDPC4_adapt2}, where the sum of the numbers of shortened and punctured bits remains constant.

Then the parties take blocks of their raw keys $\xb$ and $\yb$ of length $n-p-s$ and 
pad them with shortened and punctured symbols obtaining extended keys $\xeb$ and $\yeb$ of code block length $n$.
In Fig.~\ref{fig:methods}(a) we denote this operation as $E(\cdot, S, P)$, where $S$ and $P$ are lists of positions for shortened and punctured symbols of length $s$ and $p$, respectively.
If it is possible, parties choose $P$ using position from special list, generated in advance with untainted puncturing technique~\cite{UntP}.
Otherwise, the parties choose $P$ as well as $S$ with a synchronized PRNG.
All shortened symbols obtains zero values, while the values of the punctured bits come from the TRNG (independently on each side).
The party that modifies its raw key (in our case it is Bob) also keeps original positions of shortened and punctured symbols as $S_0$ and $P_0$.
These position are used in the final stage of the procedure.

The subsequent part of the procedure aims at reconstruction of a vector $\eb_\mathrm{dec}$, which we call an error pattern, such that 
\begin{equation}
	\xeb=\yeb+\eb_\mathrm{dec} ~ (\mathrm{mod}~2).
\end{equation}
In order to cope with this task, both Alice and Bob initialize the supposed error pattern $\eb$ as the zero vector, 
calculate the syndromes $\sbx$ and $\sby$ of their extended keys $\xeb$ and $\yeb$, and share the obtained syndromes with each other.
Then each party performs belief-propagation decoding with the relative syndrome 
\begin{equation} \label{eq:relsyndr}
	\sb=\sbx+\sby~(\mathrm{mod}~2).
\end{equation}

We use an updated belief-propagation decoding algorithm (see below), 
which returns not only a resulting decoded vector $\eb_\mathrm{dec}$ (that is ``none'' in the failure case), 
but also a set of bit positions $D$ of a fixed length $d$ which have the lowest LLR upon completion of the decoding.
In other words, the updated decoding procedure returns $d$ positions of symbols with the most uncertainty in their values.

Owing to the fact that both parties perform the same operation, they obtain the same output $\eb_\mathrm{dec}$ and $D$.
In the case of a failure ($\eb_\mathrm{dec}=\mathrm{none}$), 
the parties share the values of bits in the positions of $D$, update their supposed error pattern $\eb$ in the positions from $D$ according to received values,
\begin{equation}
	\eb[D] = \xeb[D]+\yeb[D]~(\mathrm{mod}~2),
\end{equation}
and try to perform the decoding process again by marking positions in $D$ as shortened, which is crucial for the subsequent decoding procedure.
This sequence of operations is repeated until a convergence of belief-propagation decoding.
Then Bob applies error correction according to the obtained error pattern $\ed$ by modulo 2 summation with his extended key $\yb$.
Finally, Bob excludes the symbols with initially shortened $S_0$ and punctured $P_0$ positions to obtain the corrected key $\xd$ [we denote this operation $E^{-1}$ in Fig.~\ref{fig:methods}(a)], 
and parties move to the verification step with the original raw key $\xb$ on Alice's side and its corrected version $\xd$ on Bob's one.

\setcounter{equation}{0}
\renewcommand{\theequation}{B\arabic{equation}}
\newcommand{\ab}{{\bf a}}

\section*{Appendix B: Belief propagation decoding}

We use a belief-propagation sum-product algorithm~\cite{LDPC_orig2} based one the use of LLRs, with some updates necessary for our implementation.
For a given random bit variable $X$, its LLR is defined as
\begin{equation}
	\LLR(X) \equiv \log\frac{\mathrm{Prob}(X=0)}{\mathrm{Prob}(X=1)}.
\end{equation}
One can see that the sign of LLR corresponds to the most likely value of $X$ (0 for a positive LLR and 1 for a negative one), and its absolute value exhibits the confidence level of this particular value.

The decoding algorithm is based on the representation of parity-check matrix $\Hb$ in the bipartite graph [see Fig.~\ref{fig:methods}(c)].
It consists of $n$ \emph{symbol nodes} and $m$ \emph{check nodes} that corresponds to rows and columns of parity-check matrix $\Hb$.
The $i$-th symbol node is connected by edge with the $j$-th check node if and only if the corresponding element of parity-check matrix is nonzero: $\Hb[i,j]=1$.
Process of a decoding can be described as an exchange of messages about the symbol nodes. 

We consider the decoding procedure as follows:
\begin{equation} \label{eq:synddec}
	\ed, D:=\mathrm{decode}(\sb, \eb, \Hb,\qe, S, P),
\end{equation}
where $\sb$ is the syndrome, 
$\eb$ is the vector of length $n$ that has to be corrected, 
$\Hb$ is the $m\times n$ parity-check matrix, 
$\qe$ is the estimated level of crossover probability (QBER), 
$S$ and $P$ are positions of shortened and punctured bits, 
$\ed$ is the corrected version of $\eb$, 
and $D$ is the list of positions for $d$ symbols with the lowest LLR values, where $d$ is considered as an external constant.

The workflow of the procedure is as follows.
We with a calculation of the initial LLRs for all symbol nodes.
The corresponding vector is denoted as $\rb^{(0)}$ and its elements are given by
\begin{equation}\label{eq:r0}
	\rb^{(0)}[i] := \begin{cases}
		(-1)^{\eb[i]} r_\mathrm{k}, &i\in K \\
		(-1)^{\eb[i]} r_\mathrm{s}, &i\in S \\
		0, &i\in P
	\end{cases},
\end{equation}
where $K$ consists of raw key positions, such that
\begin{equation}
	K\cup S\cup P=\{1,2,\ldots,n\}.
\end{equation}
Here, $r_\mathrm{k}$ is calculated using an estimated value of QBER, $\epse$:
\begin{equation}
	r_\mathrm{k} = \log\frac{1-q_\mathrm{est}}{q_\mathrm{est}}.
\end{equation}
The LLR value for shortened symbols $r_\mathrm{s}\gg 1$ and, in our implementation, we use $r_\mathrm{s}:=100$.
The LLR for punctured symbols is zero, as there is no information about their values since they come from an independent true RNG

The initial messages from check nodes to symbol nodes are given by the initial values of corresponding LLRs as follows:
\begin{equation}
	\Mb_{i\rightarrow j}^{(1)} := \rb^{(0)}[i].
\end{equation}
Here $i\in \Nm$ and $j\in A_i$, where $A_i$ is a set of symbol nodes connected to the $i$-th symbol node.

Check nodes from messages back to the symbol node are realized in the following way:
\begin{equation} \label{eq:Mes}
	\Mb_{i\leftarrow j}^{(k)} := 2\tanh^{-1}\left[\prod_{i'\in B_j/i}\frac{\Mb_{i'\rightarrow j}^{(k)}}{2}\right](-1)^{\sb[j]},
\end{equation}
where $j\in\Mm\equiv\{1,2,\ldots,m\}$ and $i\in B_j$, with $B_j$ being a set of symbol nodes connected to the $j$-th check node.
We note that $\Mb_{i\leftarrow j}$ does not take into account $\Mb_{i\rightarrow j}$.
Actually $\Mb_{i\leftarrow j}$ is the LLR of the $i$-th bit value based on satisfying the parity equation of the $j$-th row of the parity-check matrix $\Hb$, 
and the LLRs of all other symbol nodes taking part in this equation [see Fig.~\ref{fig:methods}(d)].

The symbol node updates its LLR using all of the messages coming from its check nodes:
\begin{equation}\label{eq:step3}
	\rb^{(k)}[i] := \rb^{(0)}[i]+\sum_{j\in A_i} \Mb_{i\leftarrow j}^{(k)}
\end{equation}
and it calculates current estimates for the bit values,
\begin{equation}
	\zb^{(k)}[i] := \begin{cases}
		0, &\rb^{(k)}[i]\geq0 \\
		1, &\rb^{(k)}[i]<0
	\end{cases}.
\end{equation}
If this estimate satisfies all parity equations, 
\begin{equation}
	\Hb\zb^{(k)}=\sb~(\mathrm{mod}~2), 
\end{equation}
then the algorithm stops and returns the decoded vector $\zb$

As a stopping criterion, we consider behavior of averaged magnitude LLRs for symbols in nonshortened positions:
\begin{equation}
	\hat{\mathbf{r}}^{(k)}:=\frac{1}{n-s}\sum_{i\in K\cup P}|\rb^{(k)}[i]|.
\end{equation}
We stop the decoding and return ``none'' as the decoded vector if, for the current step $k$, the following inequality holds:
\begin{equation}
	\hat{\mathbf{r}}^{(k)}\leq\frac{1}{\mathcal{N}}\sum_{j=k-\mathcal{N}}^{k-1}\hat{\mathbf{r}}^{(j)},
\end{equation}
where we use $\mathcal{N}:=5$.
It can be interpreted as the end of a growth trend as far as our confidence in bit values. 
The algorithm also returns a $D$ that is a list of $d$ positions of symbols which has minimal values of LLR magnitude
\begin{equation}
	D = \{i\;|\;|\rb^{(k)}[i]|\leq |\rb^{(k)}[j]| ~\forall j\notin D \}, \quad |D| = d.
\end{equation}
Otherwise, the algorithm goes to the next step.

According to the new LLRs, we update the messages from the symbol nodes to the check nodes:
\begin{equation}
\begin{split}
	\Mb_{i\rightarrow j}^{(k+1)} &:= \rb^{(0)}[i]+\sum_{j'\in A_i/j} \Mb_{i\leftarrow j'}^{(k)} \\
	&= \rb^{(k)}[i]-\Mb_{i\leftarrow j}^{(k)}
\end{split}
\end{equation}
where the counter of iterations is incremented, $k:=k+1$ [see Fig.~\ref{fig:methods}(e)], and the algorithm goes to the step with check nodes from messages back to symbol node [see Eq. (\ref{eq:step3})].

It is important to note that the most computationally expensive calculation~(\ref{eq:Mes}) can be optimized by using a technique suggested in Ref.~\cite{Decodoptim}.
We also point out that Eq.~(\ref{eq:Mes}) reveals some peculiarity regarding punctured symbols.
A zero LLR of the punctured symbol $i\in B_j$ on the first step ``deactivates'' the $j$th check node making all messages $\Mb_{i'\leftarrow j}$ ($i'\in B_j$, $i'\neq i$) to other symbol nodes to be zero.
If there are no punctured bits in $B_j/i$, then $|\Mb_{i\leftarrow j}|>0$, and the $i$th node is ``rescued'' after the first iteration and then participates in the decoding procedure. 
However, if there are at least two punctured nodes connected to a given $j$th check node, then all messages $\Mb_{i\leftarrow j}, i\in B_j$ are zero.
There still a possibility that the punctures symbols will be 'rescued' via another check nodes, but such behavior nonetheless indicates the importance of choosing a set of punctured symbols.
To avoid this situation the special technique of untainted puncturing is used~\cite{UntP}.

\setcounter{equation}{0}
\renewcommand{\theequation}{C\arabic{equation}}

\section*{Appendix C: Security of symmetric blind reconciliation}
Here we consider a security issue of the presented symmetric blind reconciliation protocol.
Specifically, we show that the amount of information leakage have to be taken into account in the privacy amplification step in accordance with Eq.\eqref{eq:fadapt} is given by
\begin{equation} \label{eq:leak}
	{\rm leak}_{\rm ec}=m-p_0+n_\mathrm{add}\;d,
\end{equation}  
where recall $m$ is the syndrome length, $p_0$ is the initial number of punctured symbols, $n_\mathrm{add}$ is the number of additional communication rounds, 
and $d$ is the number of bits disclosed in each of these rounds.

During the symmetric blind reconciliation Alice and Bob exchange 
\begin{equation}\label{eq:len}
	\mathcal{L} = 2(m+n_\mathrm{add}\;d)
\end{equation}
bits of information: each party transfers an $m$-bit syndrome and  $n_\mathrm{add}\;d$ bits in additional communication rounds.
Obviously, the information leakage could not be larger than $\mathcal{L}$.
Nevertheless, the actual leakage turns out to be lower (and given by Eq.\eqref{eq:leak}) for two reasons.
First is that the $p_0$ punctured symbols due to their inherent uncertainty decrease the leakage from each of the syndromes from $m$ to $m-p_0$; 
and second is that the factor 2 in~\eqref{eq:len} could be omitted due to the fact that each pair of parity bits 
(one transferring from Alice to Bob, and the other -- in the opposite direction) can bring only one bit of valuable information to the eavesdropper.
Next, we would like to focus on these two points.

Let us consider a symmetric blind reconciliation protocol for ${m}\times{n}$ parity-check matrix,  $(n-s_0-p_0)$-bit sifted keys, $s_0$ shortened, and $p_0$ punctured bits.
Recall that the punctured symbols are the bits with values coming from independent TRNGs on both sides. 
The consideration of the puncturing technique is greatly simplified in untainted puncturing ~\cite{UntP}, which is employed in our scheme.
In the untainted puncturing, each of the check nodes of the Tanner graph (corresponding the employed parity-check matrix) is connected with, at most, one punctured symbol.
This implies that it is possible to split al ofl the syndrome bits' positions $\{1,2,\ldots,m\}$ into the subsets $F$ and $\overline{F}$ 
where $F$ contains $p_0$ check nodes indices of those which are connected with the punctured bits, and the subset $\overline{F}\equiv \{1,2,\ldots,m\}/F$ contains all of the remaining syndrome-bit positions.
Since each of the syndrome-bit values $\sb_\xb[F]$ and $\sb_\yb[F]$ is obtained as the result of a modulo-2 summation of nonpunctured symbols and a single punctured bit (which is completely random at each side), 
the disclosure of $\sb_\xb[F]$ and $\sb_\yb[F]$ in the public channel does not bring any valuable information to the eavesdropper.
One can imagine that bits of $\sb_\xb[F]$ and $\sb_\yb[F]$ are transferred separately from all other public communications that is considered further. 
Thus, the information leakage does not exceed $2(m-p_0+n_\mathrm{add}\;d)$ bits.
We wish to remark that the case of random puncturing, where the \emph{positions} for punctured symbols are chosen with synchronised PRNGs, 
could be treated in a similar way, but in this case an additional information about the Tanner graph construction method has to be taken into account. 
One has to show that it is always possible to choose $p_0$ syndrome bits whose values are statistically independent from the $n-p_0$ values of symbols that are not punctured.

The second statement about the factor 2 in Eq.~\eqref{eq:len} is closely related to a similar question regarding to the one of the first information reconciliation protocols -- Cascade~\cite{Cascade1_orig}, 
where Alice and Bob also exchange with parity bits of some blocks of their keys in a symmetrical way.
A rigorous examination of this question related to the BB84 protocol is considered by H.-K. Lo in Ref.~\cite{LO_decoupling}.
In his work Lo considers the \emph{adaptive symmetric method for error correction} (which we will refer to simply as \emph{adaptive symmetric error correction}) --- 
the generalized information reconciliation protocol which includes Cascade and the symmetric blind reconciliation.
Next, we provide its definition.

\emph{Definition}~(adapted from Ref.~\cite{LO_decoupling}). 
Adaptive symmetric error correction is an iterative procedure applied to Alice's $n$-bit string $\xb$ and Bob's $n$-bit string $\yb$ (which is a corrupted version of $\xb$). 
where the rounds of operations given below are performed.
\begin{itemize}
	\item Round 1. Alice picks an $n$-bit string $\ab_1$, and broadcasts $\ab_1$ and the parity $p_{\xb,1} = \ab_1 \cdot \xb~\modt$. 
	Bob broadcasts the parity $p_{\yb,1} = \ab_1 \cdot \yb~\modt$. 
	Alice and Bob compute the relative parity $r_1=p_{\yb,1}-p_{\xb,1}~\modt$.
	\item Round $i$ ($i>1$). Alice picks an $i$-bit string $\ab_i$, where the choice of $\ab_i$ may depend on the strings $\{\ab_j|j<i\}$ and the relative parities $\{r_j|j<i\}$ 
	(but not on the individual parities, $\{p_{\xb,j}|j<i\}$ and $\{p_{\yb,j}|j<i\}$).
	Then Alice broadcasts $\ab_i$ and the parity $p_{\xb,i} = \ab_i \cdot \xb~\modt$. 
	Bob broadcasts the parity $p_{y,i} = \ab_i \cdot \yb~\modt$. 
	Alice and Bob compute the relative parity $r_i=p_{\yb,1}-p_{\xb,1}~\modt$.
\end{itemize}
After some number $N$ of necessary rounds, Bob corrects the errors in his string $\yb$ by applying a recovery operator 
\begin{equation}
	\zb:=\mathrm{Recover}(\yb, \{r_i\}_{i=1}^N).
\end{equation}
Ideally, $\zb = \xb$ and Alice and Bob will now share the same string $\xb$.

One can see that, in the described protocol, Alice and Bob exchange with $2N$ bits of information: $N$ bits goes from Alice to Bob, and $N$ bits -- in the opposite direction.
In Ref.~\cite{LO_decoupling} it was proven that the procedure of adaptive symmetric error correction is secure for purposes of the BB84 QKD protocol in the case where both parity bits $p_{\xb,i}$ 
and $p_{\yb,i}$ are both one-time pad encrypted with the \emph{same} secret bit.
That is, Alice and Bob exchange with the messages $p_{\xb,i}+\mathbf{w}[i]~\modt$ and $p_{\yb,i}+\mathbf{w}[i]~\modt$, 
where $\mathbf{w}$ is an $n$-bit string of the secret key possessed by Alice and Bob.
From an overall key generation balance point of view, this scenario is equivalent to the situation, where the information leakage in the adaptive symmetric error correction is equal to $N$.

Now we return to the symmetric blind reconciliation.
As it was shown, the communication of $\sb_\xb[F]$ and $\sb_\yb[F]$ can be considered separately since none of these bits bring any valuable information to the eavesdropper.
The exchange of the remaining syndrome bits $\sb_\xb[\overline{F}]$ and $\sb_\yb[\overline{F}]$, can be regarded as the first $m-p_0$ rounds of the adaptive symmetric error correction: 
for each syndrome bit, the string $\ab_i$ from the definition is the corresponding row of the employed parity-check matrix.
The syndrome decoding algorithm~\eqref{eq:synddec} either converges and allows Bob to perform a recovery of $\xb$ or returns a set of positions $D$ for the new communication round.
Note that the set $D$ is obtained as a result of the syndrome decoding executed for the relative syndrome $\sb$.
By its definition [see Eq.~\eqref{eq:relsyndr}], the relative syndrome consists of relative parities;
therefore the choice of positions for the set $D$ is in agreement with the definition of adaptive symmetric error correction.
Each of the additional communication rounds of the symmetric blind reconciliation with respect to the set $D$, 
can be regarded as $|D|=d$ of communication rounds in the definition of the adaptive symmetric error correction.
Namely, the string $\ab_i$ has all zeros except for a single one at the position $D_j$ (here, $i$ is the number of the round in the general sequence, 
and $j$ is the index of the element inside the set $D$ of length $d$).

Thus, the completion of the symmetric blind reconciliation with $n_{\rm add}$ additional rounds corresponds to the adaptive symmetric error correction with $N=m-p_0+n_{\rm add}d$ rounds, 
and the information leakage given by expression~\eqref{eq:leak}.

\end{document}